\documentclass[a4]{iopart}  
\usepackage[T1]{fontenc}
\usepackage[latin1]{inputenc}  
\usepackage{hyperref}  
\usepackage{geometry}
\usepackage{graphicx}  
\usepackage{amssymb}  
    
\date{\today}  
\def \be{\begin{equation}}
\def \bea{\begin{eqnarray}}
\def \eea{\end{eqnarray}}
\def \ee{\end{equation}}
\def \no{\nonumber}

\def \a {\alpha}

\def \Ldot{\dot L}
\def \r{{\bf r}}
\def \eps{\epsilon}

\def \Q{{\cal Q}}
\def \U{{\cal U}}
\def\lsim{\mathrel{\rlap{\lower4pt\hbox{\hskip1pt$\sim$}}
    \raise1pt\hbox{$<$}}}                
\def\gsim{\mathrel{\rlap{\lower4pt\hbox{\hskip1pt$\sim$}}
    \raise1pt\hbox{$>$}}}                

\begin{document}  
\title{Time-delay interferometry and the relativistic treatment of LISA optical links}  
\author{S. V. Dhurandhar}  
\address{ IUCAA, Postbag 4, Ganeshkind, Pune - 411 007, India. }  
  
\begin{abstract}  

  LISA is a joint space mission of the ESA and NASA for detecting low frequency gravitational radiation in the band $10^{-5} - 1$ Hz. In order to attain the requisite sensitivity for LISA, the laser frequency noise must be suppressed below the other secondary noises such as the optical path noise, acceleration noise etc. This is achieved because of the redundancy in the data, more specifically, by combining six appropriately time-delayed data streams containing fractional Doppler shifts - time delay interferometry (TDI). The orbits of the spacecraft are computed in the gravitational field of the Sun and Earth in the Newtonian framework, while the optical links are treated fully general relativistically and  thus, effects such as the Sagnac, Shapiro delay, etc. are automatically incorporated. We show that in the model of LISA that we consider here, there are symmetries inherent in the physics, which may be used effectively to suppress the residual laser frequency noise and simplify the algebraic approach to TDI.  

\end{abstract} 


\maketitle  
\section{Introduction \label{SC:1}}  
LISA - Laser Interferometric Space  Antenna - is a proposed mission which will use coherent laser beams exchanged between three identical spacecraft forming a giant (almost) equilateral triangle of side 
$5 \times 10^6$ kilometres to observe and detect low 
frequency cosmic GW \cite{RIP}. 
\par
In ground based detectors the arms are as symmetrical as possible so that the laser light 
experiences nearly identical delay in each arm of the interferometer which reduces 
 the laser frequency/phase noise at the photodetector. 
 However, in LISA, the lack of symmetry will be much larger than in terrestrial instruments.  Laser frequency noise dominates the other secondary noises, such as optical path noise, acceleration noise by 7 or 8 orders of magnitude, and must be removed if LISA is to achieve the required sensitivity of $h \sim 10^{-22}$, where $h$ is the metric perturbation caused by a gravitational wave. In LISA, six data streams arise from the exchange of laser beams between the three spacecraft  approximately 5 million km apart. These six streams produce redundancy in the data which can be used to suppress the laser frequency noise by the technique called time-delay interferometry (TDI) in which  
 the six data streams are combined with appropriate time-delays \cite{ETA}. 
\par
This work was put on a sound mathematical footing by showing that the data combinations constituted an algebraic structure; the data combinations cancelling laser frequency noise formed the {\it module of syzygies} over the polynomial ring of time-delay operators \cite{DNV}. The module was obtained - that is its generators were obtained -  for the simple case of stationary LISA in flat spacetime. These were the so-called first generation TDI. However, LISA spacecraft execute a rotational motion, the arm-lengths change with time and the background spacetime is curved, all of which affect the optical links and the time-delays. The rotation gives rise to the Sagnac effect which implies that the up-down optical links are unequal, the arm-lengths or the time-delays  change with time - flexing of arms. These effects cannot be ignored if the laser frequency noise is to be effectively cancelled. 
\par
In this paper, we compute the orbits of spacecraft in the Newtonian framework where the Earth's gravitational field is also taken into account. The base orbits we take to be Keplerian in the gravitational field of the Sun only. On these base orbits, we linearly superpose the perturbative effect of the Earth's gravitational field. We choose the Earth over Jupiter  because  (i) the Earth perturbs the Keplerian orbit in resonance, resulting in a secular growth of the perturbations and, (ii) Jupiter's effect is less than 10$\%$ of that of the Earth's on the flexing and hence not dominant. The perturbative analysis is carried out within the Clohessy-Wiltshire (CW) \cite{CW} framework. Further,  an extension of the previous algebraic approach is proposed for the general problem in which the time-delay operators in general do not commute; this leads to the second generation TDI and imperfect cancellation of laser frequency noise. However, we show that there are symmetries in the physical model which can simplify to some extent the totally non-commutative problem. These computations will be useful in the development of a LISA simulators, the LISACode for instance \cite{LISACode}.
 
\section{The spacecraft orbits in the Sun's and Earth's field}

The Keplerian orbits, the orbital motion in the gravitational field of the Sun only are chosen so that the peak to peak variation in armlengths is the least $\sim 48000$ km, see \cite{NKDV}. We summarise the results of paper 
\cite{NKDV} below.  We choose the Sun as the origin with Cartesian coordinates $\{X, Y, Z\}$ as follows:   The ecliptic plane is the $X-Y$ plane and we consider a circular reference orbit  of radius $R=$ 1 A. U. centred at the Sun.  Let $\delta_0 = 5 \a/8 $ where $\a = L_0/2R$  and $L_0 \sim 5,000,000$ km is a constant representing the nominal distance between two spacecraft of the LISA configuration. We choose the tilt of the plane of the LISA triangle to be $\delta = \pi/3 + \delta_0$ which has been shown to yield  minimum flexing of the arms.  We choose spacecraft 1 to be at its lowest point (maximum negative Z) at $t = 0$.   This means that at this point, $Y = 0$ and $X \simeq R (1-e)$. The   
orbit of the first spacecraft is an ellipse with inclination angle $\epsilon_0$,  eccentricity $e$ and satisfying the above initial condition.  
\par
From the geometry, $\epsilon_0$ and $e$ are obtained as functions of $\delta$,  
\bea  
\tan \epsilon_0 &=&  \frac{\a \sin \delta}  
{\a \cos \delta + \sin(\pi/3)} \, ,\nonumber  \\   
e &=& \left[ 1+ \frac{4}{3} \a^2   + \frac{4}{\sqrt{3}}\a   
\cos \delta \right]^{1/2} - 1 \, .  
\label{eq:eincl}  
\eea  
The equations for the orbit of spacecraft 1 are given by:  
\bea  
X_1 &=& R(\cos \psi_1 - e) \cos \epsilon_0, \no \\  
Y_1 &=& R \sqrt{1 - e^2} \sin \psi_1, \no \\  
Z_1 &=& -R(\cos \psi_1 - e) \sin \epsilon_0.   
\label{tltorb}  
\eea    
The eccentric anomaly $\psi_1$ is implicitly given in terms of $t$ by,  
\be  
\psi_1 - e \sin \psi_1 = \Omega t - \phi_0\, ,  
\label{ecc}
\ee  
where $t$ is the time and $\Omega$ is the average angular velocity and $\phi_0$ the initial phase.
The orbits of the spacecraft 2 and 3 are obtained by rotating the orbit  of spacecraft 1 by $2 \pi / 3$ and $4 \pi/3$ about the $Z-$axis; the phases $\psi_2, \psi_3$,  however, must be adjusted so that the spacecraft are at a distance $\sim L_0$ from each other. The orbital equations of spacecraft $k = 2, 3$ are:  
\bea  
X_k &=&  X_1 \cos \sigma_k- Y_1 \sin \sigma_k \, , \no \\  
Y_k &=&  X_1 \sin \sigma_k +  Y_1 \cos \sigma_k \, , \no \\  
Z_k &=& Z_1 \, ,  
\label{orbits}  
\eea  
where $\sigma_k = \left(k-1\right) \frac{2 \pi}{3}$,  
with the caveat that the $\psi_1$ is replaced by the phases $\psi_k$, where they   
are implicitly given by,  
\be  
\psi_k - e \sin \psi_k = \Omega t-\sigma_k - \phi_0.  
\label{eq:psik}  
\ee   
These are the exact (Keplerian) expressions for the orbits of the three spacecraft in the Sun's field.
\par
The Earth's field  is now included perturbatively using the CW framework. The CW frame is chosen as follows:
We take the reference particle to be orbiting in a circle of radius $R$ with constant Keplerian angular velocity 
$\Omega$. Then the transformation to the  CW frame $\{ x, y, z \}$ from the barycentric frame $\{ X, Y, Z \}$ is given by,  
\begin{eqnarray}  
x & = & \left(X-R\,\cos\Omega t\right)\,\cos\Omega t\;+\;\left(Y-R\,\sin\Omega t\right)\,  
\sin\Omega t\,,\nonumber \\  
y & = & -\left(X-R\,\cos\Omega t\right)\,\sin\Omega t\;+\;\left(Y-R\,\sin\Omega t\right)\,  
\cos\Omega t\,,\nonumber \\  
z & = & Z.\label{eq:CW}  
\end{eqnarray}   
The $x$ direction is normal and coplanar with the reference orbit, the $y$ direction is tangential and comoving, and the $z$ direction is chosen orthogonal to the orbital plane.  Linearised dynamical equations for test-particles in the   
neighbourhood of the reference particle are easily obtained. Since the frame is noninertial, Coriolis and centrifugal forces appear in addition to the tidal forces. With the help of the CW formalism, it is easy to see that to the first order in $\a$ (or equivalently $e$) there exist configurations of spacecraft so that the mutual distances between them remain constant in time. The flexing appears only when we consider second and higher order terms in $\a$. In fact in \cite{NKDV} we have shown that the second order terms describe the flexing of LISA's arms quite accurately as compared to the exact Keplerian orbits.     
\par

The CW equations for a test particle are given by:  
\bea  
\ddot{x}-2 \Omega \dot{y} - 3 \Omega^2 x & =& 0 \,  
 , \no \\  
\ddot{y} + 2 \Omega \dot{x} & = &0 \, , \no\\  
\ddot{z} +  \Omega^2 z & = &0.  
\label{gde2}  
\eea  
We choose those solutions of Eq.(\ref{gde2}) which form an equilateral triangular configuration of side $L_0$ (such solutions exist). For the $k$th spacecraft we have the  following position coordinates:
\bea
x_k &=&-\frac{1}{2} \rho_0 \cos(\Omega t - \sigma_k - \phi_0) \, , \no \\
y_k &=& \rho_0 \sin( \Omega t - \sigma_k - \phi_0) \, , \no \\
z_k &=& -\frac{\sqrt{3}}{2}\rho_0 \cos(\Omega t - \sigma_k - \phi_0) \, ,
\label{cws}
\eea 
where $\rho_0 = L_0 /\sqrt{3}$. Also at $t=0$ the initial phase of the configuration is described through $\phi_0$. In this solution, any pair of spacecraft maintain the constant distance $L_0$ between each other.
\par
LISA follows the Earth $20^{\circ}$ behind. We consider the model where the centre of the Earth leads the origin of the CW frame by $20^{\circ}$ - thus in our model, the `Earth' or the centre of force representing the Earth, follows the circular reference orbit of radius 1 A. U. Also the Earth is at a fixed position vector $\r_{\oplus} = (x_{\oplus}, y_{\oplus}, z_{\oplus})$ in the CW frame. We find that $x_{\oplus} = - R (1 - \cos 20^{\circ}) \sim - 9 \times 10^6$ km, $y_{\oplus} = R \sin 20^{\circ} \sim 5.13 \times 10^7$ km and $z_{\oplus} = 0$. The acceleration field ${\bf a}$ due to the Earth at any point $\r$ (in particular at any spacecraft) in the CW frame is given by:
\be
{\bf a} (\r) = - G M_{\oplus} \frac{\r - \r_{\oplus}}{|\r - \r_{\oplus}|^3} \,,
\ee   
where $M_{\oplus} \sim 5.97 \times 10^{24}$ kg is the mass of the Earth and 
$G = 6.67 \times 10^{-11} ~ {\rm kg}^{-1} {\rm m}^3 {\rm sec}^{-2}$ Newton's gravitational constant. 
\par
In order to write the CW equations in a convenient form we first define the small parameter $\eps$ in terms of the quantity $\omega_{\oplus}^2 = G M_{\oplus} / d_{\oplus}^3$, where  $d_{\oplus} = |\r_{\oplus}|$ is the distance of the Earth from the origin of the CW frame; $d_{\oplus} \sim 5.2 \times 10^7$ km which is more than 50 million km. So when deriving the forcing term we make the aprroximation $|\r - \r_{\oplus}| \approx d_{\oplus}$, that is, we neglect $|\r|$ compared to $d_{\oplus}$. It will turn out that the flexing due to the Earth is small so that this approximation is not unjustified. We define $\eps = \omega_{\oplus}^2 / \Omega^2 \simeq 7.16 \times 10^{-5}$ which is the just the ratio of the tidal forces due to the Earth and the Sun. The CW equations including the Earth's field take the form:
\bea  
\ddot{x}-2 \Omega \dot{y} - 3 \Omega^2 x + \eps \Omega^2 (x - x_{\oplus})& =& 0 \,, \no \\  
\ddot{y} + 2 \Omega \dot{x} + \eps \Omega^2 (y - y_{\oplus}) & = & 0 \,, \no\\  
\ddot{z} +  \Omega^2 (1 + \eps) z & = &0.  
\label{CWearth}  
\eea
Note that the compounded flexing due to the combined field of Earth and Sun is a nonlinear problem; it is infact a three body problem. We however solve this problem approximately. Assuming that both effects are small we may linearly add the flexing vectors due to the Sun and Earth; that is, add the perturbative solutions obtained from Eqs.(\ref{gde2}) and (\ref{CWearth}); the nonlinearities appear at higher orders in $\a$ and $\eps$. These would modify the flexing but we may neglect this effect because of the smallness. We find that the flexing produced by the Earth is of the order of 1 or 2 m/sec upto the third year, just about 40 $\%$ of that due to the Sun. But, as shown in \cite{NKDV} the flexing produced by the Sun's octupole field is nearly exact to that produced by the Keplerian orbits. Thus we may do better by just adding the flexing vector produced by the Earth to the Keplerian orbit of the relevant spacecraft. 
\par   
We then seek perturbative solutions to Eq. (\ref{CWearth}) to the first order in $\eps$. We write, 
$x = x_0 + \eps x_1, y = y_0 + \eps y_1, z = z_0 + \eps z_1$ where $x_0, y_0, z_0$ are solutions at the zeroth order given by Eq.(\ref{cws}). We put $\sigma_k = 0$ (or equivalently include it in $\phi_0$) in these solutions for simplifying the algebra. 
\par
With the initial conditions: $x_1 = y_1 = z_1 = {\dot x_1} = {\dot y_1} = {\dot z_1} = 0$ at $t = 0$, we have the results:
\bea
x_1 &=& - \rho_{\oplus} \cos (\Omega t - \phi_{\oplus}) + x_{\oplus} + 2 y_{\oplus} \Omega t - 2 \rho_0 \cos \phi_0 + \frac{5}{4} \rho_0 \Omega t \sin (\Omega t - \phi_0) \,, \no \\
y_1 &=& 2 \rho_{\oplus} [\sin (\Omega t - \phi_{\oplus}) + \sin \phi_{\oplus}] - \frac{3}{2} \rho_0 [\sin (\Omega t - \phi_0) + \sin \phi_0] \no \\
&& + \frac{5}{2} \rho_0 \Omega t \cos (\Omega t - \phi_0) - \Omega t (2 x_{\oplus} - 3 \rho_0 \cos \phi_0) - \frac{3}{2} \Omega^2 t^2 y_{\oplus} \,,
\label{solnxy}
\eea
where,
\bea
\rho_{\oplus}^2 &=& (x_{\oplus} - 2 \rho_0 \cos \phi_0)^2 + (2 y_{\oplus} - \frac{5}{4} \rho_0 \sin \phi_0)^2  \,, \no \\
\tan \phi_{\oplus} &=& \frac{2 y_{\oplus} - \frac{5}{4} \rho_0 \sin \phi_0}{x_{\oplus} - 2 \rho_0 \cos \phi_0} \,.
\eea
The $z$ equation can be exactly integrated and used directly to obtain the flexing. However, we can also expand this solution to the first order in $\eps$ and the result is:
\be
z_1 = \frac{\sqrt{3}}{4} \rho_0 [\Omega t \sin \Omega t \cos \phi_0 - (\Omega t \cos \Omega t - \sin \Omega t) \sin \phi_0] \,.
\label{solnz}
\ee
As argued before, we add the perturbation given by $\eps \r_1 = \eps (x_1, y_1, z_1)$ to the Keplerian orbit of each spacecraft. Next we compute the optical links. 

\section{The six optical links}

\begin{figure}[h!]  
\centering  
\includegraphics[width = 0.6\textwidth]{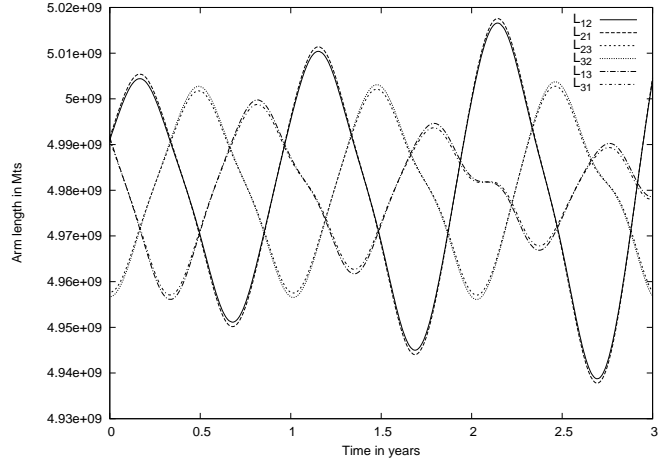}  
\caption{The figure shows the variation in the six optical links of the LISA model for three  years for $\phi_0 = 0$. The lengths are given in metres.}  
\label{optlink}  
\end{figure}

\begin{figure}[h!]  
\centering  
\includegraphics[width = 0.6\textwidth]{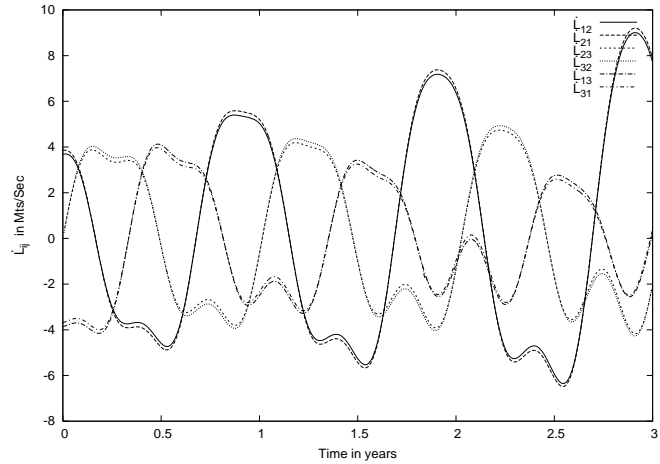}  
\caption{The rate of change of armlengths for the six links is shown in units of m/sec for $\phi_0 = 0$. This rate of change is less than 6 m/sec upto the second year and increases to a maximum of about  8 m/sec in the third year.}  
\label{dplr}  
\end{figure}

The time-delay that is required for the TDI operators needs to be known very accurately - at least to 1 part in $10^8$, that is, to about few metres - for the laser frequency noise to be suppressed. In order to guarantee such level of accuracy, we numerically compute the optical links or the time-delay. This approach is guaranteed to give the desired accuracy or even better accuracy than what is required. We numerically integrate the null geodesics followed by the laser ray  emitted by one spacecraft and received by the other. This computation is performed in the barycentric frame, and taking into account the fact that the spacetime is curved by the Sun's mass only (the Earth's contribution is about 5 orders of magnitude less). The computation here is further complicated by the fact that the spacecraft are moving in this frame of reference and the photon emitted from one spacecraft must be received by the other spacecraft. We use the Runga-Kutta numerical scheme to integrate the differential equations describing the null geodesics. But since the end point of the photon trajectory is not known apriori, an iterative scheme must be devised for adjusting the parameters of the null geodesic, in order that the worldlines of the  photon and the receiving spacecraft intersect. We have devised such a scheme based on the difference vector between the photon position vector and receiving spacecraft position vector. The six optical links $L_{ij}$ have thus been numerically computed with sufficient accuracy required for TDI. The code gives results accurate to better than 10 metres - most of the time better than $10^{-2}$ metres - except in a window of about half an hour when the error exceeds this value and becomes unacceptably large. We display the results in Figures \ref{optlink} and \ref{dplr} for $\phi_0 = 0$. More details may be found in \cite{DVN08}.
\par
Figure \ref{optlink} shows all the six optical links in the combined field of the Sun and Earth. 
We also need to estimate the variation in armlength which is important for the TDI analysis to follow. Figure \ref{dplr} shows the rate of change of the six optical links as a function of time over a period of three years.
We find that in the optimised model of LISA configuration, this rate of change is less than 4 m/sec. if we just consider the Sun's field. Including the Earth's field the flexing still remains $\lsim 6$ m/sec in the first two years and increases to $\lsim 8$ m/sec in the third year. Earlier estimates were $\sim 10$ m/sec. These numerical estimates are most crucial for their effect on residual laser frequency noise in the TDI. 

\section{Time-delay interferometry}

 In order to cancel the laser frequency noise, time-delayed data streams are added together in which an appropriate set of time-delays are chosen. In general the time-delays are multiples of the photon transit time between pairs of spacecraft. In \cite{DNV} a scheme based on modules over commutative rings was given where the module of data combinations cancelling the laser noise was constructed. This fully cancels the laser frequency noise for stationary LISA. There are only three delay operators corresponding to the three armlengths and the time-delay operators commute. This scheme can be straight forwardly extended to moving LISA \cite{NV}, where, now because of Sagnac effect, the up and down optical links have different armlengths but the  armlengths are still  constant in time. In this case, there are six delay operators corresponding to the six optical links and they commute. These are the modified but still first generation TDI.  However, for LISA the armlengths do change with time - flexing of the arms - and the first generation TDI modified or otherwise do not cancel of the laser frequency noise sufficiently. 

\subsection{Equations for the polynomial vector cancelling laser frequency noise}

We follow the notation and conventions of \cite{NV} and \cite{DNV} which are the simplest for our purpose. The six links are denoted by $U^i, V^i , i = 1, 2, 3$. The time-delay operator for the link $U^2$ from S/C 1 to S/C 2 or $1 \longrightarrow 2$ is denoted by $x$ in \cite{NV} and so on in a cyclic fashion. The delay operators in the other sense are denoted by $l, m, n$;  the link $-V^1$ from $2 \longrightarrow 1$ by $l$ and similarly the links $V^2, V^3$ are defined through cyclic permutation. 
\par
Let $C_i (t) = \Delta \nu_i (t)/\nu_0$ represent the laser frequency noise in S/C $i$. Let $j$ be the delay operator corresponding to the variable armlength $L_j (t)$, i.e. $ j C_i (t) = C_i (t - L_j (t))$. Then we have,
\bea
U^1 &=& C_1 - z C_3 \,, \no \\
V^1 &=& l C_2 - C_1 \,.
\eea
The other links in terms of $C_i (t)$ are obtained by cyclic permutations. Also in the $U^i, V^i$ we have not included contributions from the secondary noises, gravitational wave signal etc. since here our aim is to deal with laser frequency noise only. Any observable $X$ is written as:
\be   
X = p_i V^i + q_i U^i \,,
\ee
where $p_i, q_i, i = 1,2,3$ are polynomials in the variables $x,y,z,l,m,n$. Thus $X$ is specified by giving the six tuple polynomial vector $(p_i, q_i)$.  Writing out the $(V_i, U_i)$ in terms of the laser noises $C_i (t)$, and in order that the laser frequency noise cancel for arbitrary $C_i (t)$,  the polynomials $(p_i, q_i)$ must satisfy the equations:
\bea
p_1 - q_1 +  q_2 x -  p_3 n &=& 0, \no \\
p_2 - q_2 +  q_3 y -  p_1  l &=& 0, \no \\
p_3 - q_3 +  q_1 z -  p_2 m &=& 0.
\label{lneq}
\eea
The solutions to these equations as realised in earlier works are important, because they consist of polynomials with lowest possible degrees and  thus are simple. Since these are linear equations they define a homomorphism of modules and the solutions themselves form a module - the {\it module of syzygies} over the polynomial ring $\Q(x, y, z, l, m, n)$, where $\Q$ is the field of rational numbers and $x, y, z, l, m, n$ play the role of indeterminates. In general,  the variables (operators) $x,y,z,l,m,n$ do not commute and hence the order of the variables is important. However, if we assume in a simple model that the arms do not  flex, then the operators commute, and the generators of the module have been found  via Gr\"obner basis methods \cite{DNV,NV}.  

However, when the arms flex, the operators no longer commute. If we operate on $C(t)$ with operators $j$ and $k$ in different orders, it is easily seen that $ jk \neq kj$. A combinatorial approach has been adopted in \cite{Vallis} to deal with the totally non-commutative case. However, our aim here is to estimate the level of the non-commutativity of these operators in the context of our LISA model and use the symmetries to simplify the algebraic approach. 

\subsection{The algebraic approach and symmetries}

The level of non-commutativity can be found by computing commutators which occur in several of the well known TDI observables like the Michelson, Sagnac etc. We find that given our model of LISA, we require to go only upto the first order in ${\dot L}$; we find for our model $\ddot L \sim 10^{-6}$ metres/sec$^2$ and thus even if one considers say 6 successive optical paths, that is, about $\Delta t \sim 100$ seconds of light travel time, $\Delta t^2 {\ddot L} \sim 10^{-2}$ metres. This is well below few metres and thus can be neglected in the residual laser noise computation. Moreover, ${\dot L}^2$ terms (and higher order) can be dropped since they are of the order of $\lsim 10^{-15}$ (they come with a factor $1/c^2$) which is much smaller than 1 part in $10^8$. The calculations which follow neglect these terms. 

Applying the operators twice in succession and dropping higher order terms as explained above,
\bea
k_2 k_1 C &=& C(t - L_{k_1} (t - L_{k_2}) - L_{k_2}) \,, \no \\
    &\approx&  C(t -L_{k_1} - L_{k_2}) + L_{k_2} {\dot L}_{k_1} {\dot C} (t -L_{k_1} - L_{k_2}) \,.
\eea
The above formula can be easily generalised by induction to $n$ operators.
\par
We now turn to the commutators of the operators. The term in $C$ cancels out; only the ${\dot C}$ term remains. We list below a few of the  commutators:
\bea
jk-kj &=& L_j \Ldot_k - L_k \Ldot_j \,, \no \\
lmjk - jklm &=& (L_l + L_m)(\Ldot_j + \Ldot_k) - (L_j + L_k)(\Ldot_l + \Ldot_m) \,, \no \\
lmnxyz - xyzlmn &=& (L_l + L_m + L_n)(\Ldot_x + \Ldot_y + \Ldot_z) \\
                && - (L_x + L_y + L_z)(\Ldot_l + \Ldot_m + \Ldot_n) \,.
\label{commut}
\eea
We observe the following approximate symmetries in our model:
\be
\Ldot_x \approx \Ldot_l,~~ \Ldot_y \approx \Ldot_m, ~~ \Ldot_z \approx \Ldot_n \,,
\label{comm}
\ee
which also implies (this combination occurs in the Sagnac observables),
\be
\Ldot_x + \Ldot_y + \Ldot_z \approx \Ldot_l + \Ldot_m + \Ldot_n \,.
\label{cyclic}
\ee
Infact in our model, $|(\Ldot_x + \Ldot_y + \Ldot_z) - (\Ldot_l + \Ldot_m + \Ldot_n)| \lsim 1$ m/sec and $|\Ldot_x - \Ldot_l| \lsim 0.8$ m/sec upto the first three years in our model. The same is essentially true for the pairs of links $y, m$ and $z, n$. Thus these pairs of operators essentially commute. Thus, we are not dealing with a set of totally non-commuting variables, but with an intermediate case.
\par
In addition to these approximate symmetries there are other commutators which vanish `identically' (after dropping terms in $\Ldot^2$ and ${\ddot L}$ and higher order). It can be easily verified that commutators of the form, 
$[x_1 x_2 ... x_n, y_1 y_2 ... y_n], ~ n \geq 2$ vanish identically, when the $y_1, y_2, ..., y_n$ are permutation of the operators $x_1, x_2, ..., x_n$, and the $x_k$ for a given $k$ represents one of the delay operators  $x, y, z, l, m, n$. Thus the original non-commutative ring $\Q(x, y, z, l, m, n)$ can be quotiented by the ideal $\U$ which is generated by the vanishing and approximately vanishing commutators:  $\U = \{[x, l], [y, m], [z, n]; [xy, yx], ...\}$. The quotient ring 
$\Q/\U$ is clearly much smaller and simpler and the solution  to Eq. (\ref{lneq}) is sought  for polynomials in this quotient ring. The solution set of polynomial vectors $(p_i, q_i)$ still form a module over $\Q/\U$. The future goal is to `construct' this module. 
 
\subsection{Residual laser frequency noise in the Sagnac observable and symmetries} 

 By the time LISA flies the expectations are for the laser frequency noise estimate to reduce to say ${\widetilde{\Delta \nu}} \sim 10 {\rm Hz}/\sqrt{\rm Hz}$. If we divide this number by the laser frequency $\nu_0 \sim 3 \times 10^{14}$ Hz, we obtain the noise estimate in the fractional Doppler shift $C$ with the power spectral density (PSD):
\be
S_C (f) = \langle |{\tilde C} (f)|^2 \rangle \sim 10^{-27}~~~ {\rm Hz}^{-1} \,,
\label{lsr_noise}
\ee
where ${\tilde C} (f)$ is the Fourier transform of $C(t)$. Then by differentiating $C$, the PSD of the random variable ${\dot C}$ is just  $S_{\dot C} (f) = 4 \pi^2 f^2 S_C (f)$ Hz.
\par
\begin{figure}[h]  
\centering  
\includegraphics[width = 0.5\textwidth, angle=-90]{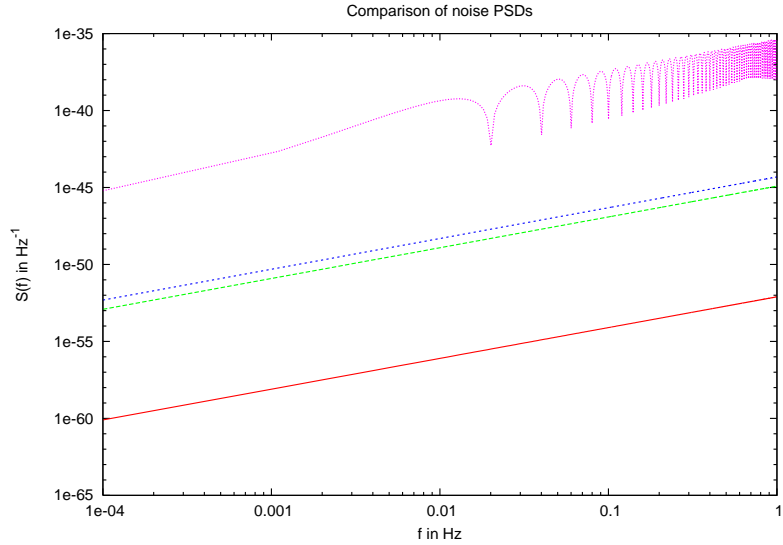}  
\caption{The `top' curve shows the PSD $S_{\a} (f)$ of the secondary noises. The straight lines are the PSDs of the residual noise at three epochs chosen an year apart. Clearly the residual laser noise is adequately below the secondary noises.}  
\label{sagnac}  
\end{figure}

The modified Sagnac first generation TDI observable $\alpha$ is given by the polynomial vector in the form $(p_i, q_i)$ by:
\be
\alpha = (\kappa, \kappa l, \kappa lm, \eta, \eta zy, \eta z) \,,
\label{sgnc}
\ee 
where $\kappa = 1 - zyx$ and $\eta = 1 - lmn$. If the variables $x,y,z,l,m,n$ commute then the laser frequency noise is fully cancelled. However, if they do not commute, there is a residual term.  It can be computed as:
\be
\Delta C = \a_1 C_1 + \a_2 C_2 + \a_3 C_3 \,.
\ee 
We find that $\a_2 = \a_3 \equiv 0$ and $\a_1 = [zyx, lmn]$ and so by Eq. (\ref{commut}): 
\be
\Delta t (t) = \frac{1}{c^2}[(L_x + L_y + L_z)(\Ldot_l + \Ldot_m + \Ldot_n) - (L_l + L_m + L_n)(\Ldot_x + \Ldot_y + \Ldot_z) ] \,,
\ee
and thus $\Delta C = \Delta t {\dot C_1}$. Because the $L_k$ vary during the course of an year the $\Delta t$ also varies during the year and so also the amplitude of the random variable $\Delta C$. Thus the PSD of $\Delta C$ is:
\be
S_{\Delta C} (f;t) = 4 \pi^2 \Delta t(t)^2 f^2 S_C(f) \,.
\label{rsdl}
\ee
This noise must be compared with the secondary noise \cite{RIP}. However, because we are considering the modified TDI Eq. (\ref{sgnc}), there are extra factors $\kappa$ and $\eta$ which do not appear in the corresponding first generation TDI. These factors introduce an additional multiplicative factor, namely, $4 \sin^2 (3 \pi f L_0)$ in the secondary noise PSD which leaves the SNR unchanged but must be considered when it is compared with the residual laser frequency noise given in Eq. (\ref{rsdl}). Thus,
\be
S_{\a} (f) = 4 \sin^2 (3 \pi f L_0)\{[8 \sin^2 3 \pi f L_0 + 16 \sin^2 \pi f L_0] S_{acc} + 6 S_{opt}\}
\ee
where $S_{acc} = 2.5 \times 10^{-48} (f/1 {\rm Hz})^{-2} {\rm Hz}^{-1}$ and 
$S_{opt} = 1.8 \times 10^{-37} (f/1 {\rm Hz})^2 {\rm Hz}^{-1}$.
In the Figure \ref{sagnac} we plot $S_{\a} (f)$ and $S_{\Delta C} (f;t)$ at three epochs an year apart. 

We see that, clearly the residual laser frequency noise is few orders of magnitude below the secondary noises. Since the other Sagnac variables $\beta$ and $\gamma$ are obtained by cyclic permutations of the spacecraft, the residual laser noise is similarly suppressed in them. The basic reason for this remarkable cancellation is the symmetry inherent in the physics. Note that $\a_1 \in \U$ and hence $\a$ is an element of the module we are seeking.

\section{Concluding remarks}

 We have computed in the Newtonian framework the spacecraft orbits in the combined field of the Sun and Earth and from this deduced the flexing of the arms of LISA by choosing the model which gave minimum flexing when only the Sun's field was taken into account. Now the flexing is no more periodic as was the case when only the Sun's field was considered. We have ignored the effect of Jupiter because we believe this effect to be not so dominant as that of the Earth. Writing the tidal parameter for Jupiter, $\eps_J = G M_J/d_J^3$, similar to $\eps$ of Earth, where, $M_J \approx 2 \times 10^{27}$ kg is the mass of Jupiter and $d_J$, the distance from LISA to Jupiter, which we take on the average to be $\sim 5$ A. U., we find $\eps_J/\eps \sim 0.09$.  Moreover, Jupiter has its own periodicity pertaining to its orbit and therefore will not be in resonance as was the case with the Earth, and thus there will be no secular effect. Thus we do not expect the effect of Jupiter on flexing to dominate. Note that these results are valid so long as we can neglect the nonlinearities arising from higher order terms in $\eps$ and $\a$.  
\par
We have computed the residual laser frequency noise in one of the  important TDI variables, namely, the Sagnac. The residual noise is satisfactorily suppressed because of the symmetry. In other variables such as the Michelson this is not true and higher degree polynomials will be required. The algebraic approach outlined above seems promising. 
\par
Our model of LISA is optimal (minimal flexing of arms) only in the Sun's field. Clearly this opens up the question of seeking an optimal model for the LISA configuration in the field of the Sun, Earth, Jupiter and other planets which will minimise the flexing of the arms and therefore the residual laser frequency noise in the modified first generation TDI. 
We finally remark that our computations here may be useful in the development of a LISA simulator. 
\ack

The author would like to thank the Indo-French Centre for the Promotion of Advanced Research (IFCPAR) project no. 3504-1 under which this work has been carried out. This work is in collaboration with J-Y Vinet and R. Nayak.

\end{document}